\documentclass[aps,prc,preprint,tightenlines,groupedaddress,nofootinbib,showpacs,preprintnumbers,amsmath,amssymb,superscriptaddress]{revtex4} 
\usepackage{graphicx}
\usepackage{dcolumn}
\usepackage{bm}
\usepackage{amsmath} 
 
\begin{document} 
 
\title{Insensitivity of the elastic proton-nucleus reaction \\
       to the neutron radius of ${}^{208}$Pb} 

\author{J. Piekarewicz}
\email{jorgep@csit.fsu.edu}
\affiliation{Department of Physics, Florida State University, 
             Tallahassee FL 32306}

\author{S.P Weppner}
\email{weppnesp@eckerd.edu}
\affiliation{Natural Sciences, Eckerd College, 
             St. Petersburg FL 33711}

\date{\today} 
 
\begin{abstract} 
The sensitivity---or rather insensitivity---of the elastic
proton-nucleus reaction to the neutron radius of ${}^{208}$Pb is
investigated using a non-relativistic impulse-approximation
approach. The energy region ($T_{\rm lab}\!=\!500$~MeV and $T_{\rm
lab}\!=\!800$~MeV) is selected so that the impulse approximation may
be safely assumed. Therefore, only free nucleon-nucleon scattering
data are used as input for the optical potential.  Further, the
optical potential includes proton and neutron ground-state densities
that are generated from accurately-calibrated models. Even so, these
models yield a wide range of values (from $0.13$~fm to $0.28$~fm) for
the poorly known neutron skin thickness in ${}^{208}$Pb.  An excellent
description of the experimental cross section is obtained with all
neutron densities.  We have invoked analytic insights developed within
the eikonal approximation to understand the insensitivity of the
differential cross section to the various neutron densities. As the
diffractive oscillations of the cross sections are controlled by the
matter radius of the nucleus, the large spread in the neutron skin
among the various models gets diluted into a mere 1.5\% difference in
the matter radius. This renders ineffective the elastic reaction as a
precision tool for the measurement of neutron radii.
\end{abstract} 
 
 
\maketitle 

\section{Introduction} 
\label{sec:intro} 
The neutron radius of a heavy nucleus---such as ${}^{208}$Pb---is a
fundamental nuclear structure quantity that remains elusive. This is
in striking contrast to the proton radius that is known with exquisite
accuracy for a variety of nuclei throughout the periodic
table~\cite{Fri95_NDT60}.  For example, the charge radius of
${}^{208}$Pb is known with a 0.01\% accuracy ($R_{\rm
ch}\!=\!5.5013(7)$~fm). The origin of the mismatch in our knowledge
of such similar quantities is the availability for over half a
century of high-quality electron beams to accurately map the nuclear
charge distribution~\cite{Hof53_PR91}. Instead, mapping the neutron
distribution has relied on strongly-interacting probes that suffer
from controversial uncertainties in the reaction
mechanism~\cite{Ray85_PRC31,Ray92_PR212}, thereby hindering the
extraction of such fundamental observable.

Parity violating electron nucleus scattering offers an attractive
alternative to the hadronic program.  The Parity Radius Experiment
(PREX) at the Jefferson Laboratory aims to measure the neutron radius
of $^{208}$Pb accurately (to within $0.05$~fm) and model independently
via parity-violating electron
scattering~\cite{Hor01_PRC63,Mic02_JLAB003}.  Parity violation is
sensitive to the neutron density because the $Z^0$ boson couples
primarily to neutrons, as its vector coupling to the proton is
suppressed by the weak mixing angle
($1\!-\!4\sin^{2}\theta_{W}\!\approx\!0.076$).  The parity-violating
asymmetry, while small, can be interpreted with as much confidence as
electromagnetic scattering experiments.  PREX should provide a
unique observational constraint on the thickness of the neutron skin
of a heavy nucleus.

While measuring the neutron radius of a heavy nucleus should be reason
enough to endorse the experiment, the search for this fundamental
quantity has been recently stimulated by theoretical evidence that a
precise measurement of the neutron radius in ${}^{208}$Pb could place
an important new constraint on the neutron-matter equation of
state~\cite{Bro00_PRL85}. Moreover, it has been also suggested that an
accurate measurement of the neutron radius in ${}^{208}$Pb will impact
strongly on a host of neutron-star observables, such as on their 
structure, composition, and cooling
mechanism~\cite{Hor01_PRL86,Hor01_PRC64,Hor02_PRC66,Car03_APJ593,Ste05_PR411}.
Prompted by these new developments, several new analysis of
proton-nucleus elastic scattering have been
reported~\cite{Kar02_PRC65,Cla03_PRC67,Dup05_xxx}.

In Ref.~\cite{Kar02_PRC65} an analysis of both proton-nucleus ($pA$)
and neutron-nucleus ($nA$) elastic scattering suggested a value for
the neutron skin of ${}^{208}$Pb---defined as the difference between
its neutron ($R_{n}$) and proton ($R_{p}$) root-mean-square 
radii---of $\Delta R\!\equiv\! R_{n}\!-\!R_{p}\!\sim\!0.17$~fm. The 
model employs a sophisticated ``g-folding'' optical
potential~\cite{Amo00_ANP25} that incorporates proton densities
constrained from electron scattering and several model neutron
densities. The neutron densities employed were generated from either
accurately calibrated Skyrme parameterizations or from simple
harmonic-oscillator models. While the harmonic-oscillator parameters
were adjusted to reproduce some average properties of ${}^{208}$Pb,
the Skyrme model parameters were generated from a bona fide fit
to a large body of ground-state data.  Calculations were performed at
laboratory kinetic energies of $T_{\rm lab}\!=\!40, 65$, and $200$~MeV
for ${}^{40}$Ca and ${}^{208}$Pb.  While the proton-${}^{208}$Pb
differential cross section at $T_{\rm lab}\!=\!200$~MeV shows clear
differences at large scattering angles between the Skyrme and
harmonic-oscillator densities (with the former being clearly favored) 
the differences among the accurately calibrated sets appears to be
marginal.

The value of $\Delta R\!\sim\!0.17$~fm proposed in
Ref.~\cite{Kar02_PRC65} is consistent with a very recent analysis that
suggests a neutron skin in the range 
$\Delta R\!\sim\!0.13\!-\!0.17$~fm~\cite{Dup05_xxx}.  While this 
$T_{\rm lab}\!=\!121$~MeV calculation is also based on a g-folding 
approach, it employs, in addition to the SKM* interaction responsible 
for the $\Delta R\!\sim\!0.17$~fm value, a neutron density generated 
in a random-phase-approximation (RPA) that yields the smaller value 
of $0.13$~fm for the neutron skin.

Finally, the global analysis of medium-energy 
($T_{\rm lab}\!=\!500\!-\!1040$~MeV) $pA$ elastic-scattering data of
Ref.~\cite{Cla03_PRC67} reports a skin thickness in ${}^{208}$Pb of
$\Delta R\!\sim\!0.083\!-\!0.111$~fm, slightly smaller than the
previous two analyzes. The extraction of neutron densities, not only
for the case of ${}^{208}$Pb but also for ${}^{40}$Ca and ${}^{48}$Ca,
is based on a global Dirac-phenomenological approach.  The merit of
the approach is (at least) twofold. First, the analysis is limited to
an energy region that is well described by the impulse
approximation. Thus, the approach relies on existent experimental
on-shell nucleon-nucleon ($NN$) data.  This avoids modeling important
and often complicated medium modifications to the $NN$ interaction.
Second, the various parameters characterizing the neutron density are
determined from a global fit to a large body of scattering observables
(differential cross sections and spin observables) at a variety of
projectile energies. What emerges from this global fit is the optimal
set of parameters characterizing the neutron densities. In stark
contrast to an earlier analysis by Ray~\cite{Ray92_PR212}, this
approach produces energy-independent neutron radii.

Nuclear giant resonances have also been used recently to place some
constraints on the neutron radius of ${}^{208}$Pb~\cite{Pie04_PRC69}.
In order to simultaneously reproduce the isovector giant dipole
resonance in ${}^{208}$Pb as well as the isoscalar giant-monopole
resonance in both ${}^{208}$Pb and ${}^{90}$Zr, a neutron skin in
${}^{208}$Pb of $\Delta R\!\alt\!0.21$~fm was
required~\cite{Tod05_PRLxx}. While this value is significantly lower
than the ones obtained with earlier relativistic parameter
sets~\cite{Lal97_PRC55}, it remains higher relative to the values
obtained from the $NA$ scattering analyzes described above.

In our effort to test the sensitivity of nucleon-nucleus scattering 
to the neutron radius of ${}^{208}$Pb, we have modified the isovector
parameters of the model so that the neutron radius of ${}^{208}$Pb may
be adjusted---while maintaining the proton radius fixed at its
experimental value.  Our numerical calculations of the differential
cross section are in very good agreement with experimental data and
show no clear distinction among the various nuclear-structure
models. To elucidate the insensitivity of the differential cross
section to the neutron density, we have invoked analytic insights
developed within the eikonal
approximation~\cite{Ama80_PRC21,Ama85_ANP15}. Such analyzes reveal
that while the filling of the diffractive oscillations germane to the
differential cross section are determined by the underlying dynamics,
the diffractive oscillations and exponential falloff of the cross
sections are dominated by the nuclear geometry---with the radius
controlling the oscillations and the diffuseness parameter the
exponential falloff. Thus the nuclear radius leaves its imprint in the
diffractive oscillations of the differential cross section. However,
at the medium energies considered here, the cross section is dominated
by the isoscalar (or matter) density. Hence, the factor of two
difference in the model dependence of the neutron skin translates into
a minute 1.5\% difference in the matter radius that can not be
resolved. We conclude that the elastic reaction may not serve as a
precision tool for the measurement of neutron radii.

The manuscript has been organized as follows. In
Sec.~\ref{sec:formalism} we present the formalism employed in the
calculations. This section includes a subsection that introduces the
various nuclear-structure models employed to generate proton and
neutron densities. An additional subsection discusses how these
nuclear densities are incorporated into the various treatments of the
optical potential. In Sec.~\ref{sec:results} we present results for
the differential cross section for the scattering of 
$T_{\rm lab}\!=\!500$ and $800$~MeV protons from ${}^{208}$Pb. Finally, 
we offer our conclusions in Sec.~\ref{sec:conclusions}.

\section{Formalism} 
\label{sec:formalism}
 
This section starts with a discussion of the nuclear-structure
model employed in the calculations. In particular, it includes four
parameterizations that yield neutron skins in ${}^{208}$Pb ranging 
from $\Delta R\!=\!0.13$~fm to 
     $\Delta R\!=\!0.28$~fm, while keeping the charge
radius fixed at its experimental value. Next, we discuss briefly
how these various densities get incorporated into the treatment of
the optical potential.

\subsection{Nuclear-Structure Models} 
\label{sec:structure}

The nuclear densities employed in this contribution are derived from
an effective field theoretical model that provides an accurate 
description of the ground-state properties of finite nuclei. The model 
is based on an interacting Lagrangian that has the following
form~\cite{Mue96_NPA606,Hor01_PRL86,Hor01_PRC64}:
\begin{eqnarray}
&&
{\cal L}_{\rm int} =
\bar\psi \left[g_{\rm s}\phi   \!-\!
         \left(g_{\rm v}V_\mu  \!+\!
    \frac{g_{\rho}}{2}{\mbox{\boldmath $\tau$}}\cdot{\bf b}_{\mu}
                               \!+\!
    \frac{e}{2}(1\!+\!\tau_{3})A_{\mu}\right)\gamma^{\mu}
         \right]\psi \nonumber \\
                   && -
    \frac{\kappa}{3!} (g_{\rm s}\phi)^3 \!-\!
    \frac{\lambda}{4!}(g_{\rm s}\phi)^4 \!+\!
    \frac{\zeta}{4!}
    \Big(g_{\rm v}^2 V_{\mu}V^\mu\Big)^2 \!+\!
    \Lambda_{\rm v}
    \Big(g_{\rho}^{2}\,{\bf b}_{\mu}\cdot{\bf b}^{\mu}\Big)
    \Big(g_{\rm v}^2V_{\mu}V^\mu\Big) \;.
\end{eqnarray}
The Lagrangian density includes Yukawa couplings of the nucleon field
($\psi$) to isoscalar scalar ($\phi$) and vector ($V^{\mu}$) meson 
fields. This ``minimal'' model---introduced by Walecka more than 30 
years ago~\cite{Wal74_Ann74}---provides a natural explanation for the
saturation of symmetric nuclear matter. To describe asymmetric matter
the model was extended by Serot to the isovector sector by introducing
an isovector-vector meson (${\bf b}^{\mu}$). Using these three meson
fields---plus the photon---Horowitz and Serot performed the first
mean-field calculations of the ground-state properties of finite
nuclei~\cite{Hor81_NPA368}. Yet the model had to be supplemented by
nonlinear scalar-meson interactions ($\kappa$ and $\lambda$) to soften
the equation of state of symmetric nuclear matter, thereby obtaining
compression moduli consistent with experiment.  The mixed
isoscalar-isovector coupling ($\Lambda_{\rm v}$) is used to modify the
density dependence of the symmetry energy and thus tune the neutron
radius of ${}^{208}$Pb without affecting other ground-state
observables~\cite{Hor01_PRL86,Hor01_PRC64}. This freedom is exploited
here to generate ground-state densities having neutron skins in the
range of $\Delta R\!=\!0.13\!-\!0.28$~fm. Finally, the quartic
vector-meson coupling ($\zeta$) impacts the high-density
component of the equation of state and is therefore weakly sensitive
to the ground-state properties of finite nuclei~\cite{Mue96_NPA606}.
The existence of powerful new telescopes studying neutron stars at a
variety of wavelengths will provide important constraints on the
poorly-known high-density component of the equation of state.

\begin{table*}
\begin{tabular}{|l||c|c|c|c|c|c|c|c|}
 \hline
 Model & $m_{\rm s}$  & $g_{\rm s}^2$ & $g_{\rm v}^2$ & $g_{\rho}^2$
       & $\kappa$ & $\lambda$ & $\zeta$ & $\Lambda_{\rm v}$\\
 \hline
 \hline
 NL3      & 508.194 &  104.387 & 165.585 &  79.600
          &   3.860 & $-$0.016 &   0.000 &   0.000 \\
 \hline
 FSUGold  & 491.500 &   112.200 & 204.547 & 138.470
          &   1.420 &  $+$0.024 &   0.060 &   0.030 \\
 \hline
 FSUGold4 & 491.000 &   112.200 & 204.547 & 182.622
          &   1.420 &  $+$0.024 &   0.060 &   0.040 \\
 \hline
 FSUGold5 & 490.250 &   112.200 & 204.547 & 268.103
          &   1.420 &  $+$0.024 &   0.060 &   0.050 \\
 \hline
\end{tabular}
\caption{Model parameters used in the calculations. The parameter
$\kappa$ and the inverse scalar range $m_{\rm s}$ are given in MeV.
The nucleon, omega, and rho masses are kept fixed at $M\!=\!939$~MeV,
$m_{\omega}\!=\!782.5$~MeV, and $m_{\rho}\!=\!763$~MeV, respectively.}
\label{Table0}
\end{table*}

\begin{figure}[ht]
\vspace{0.50in}
\includegraphics[width=5in,angle=0]{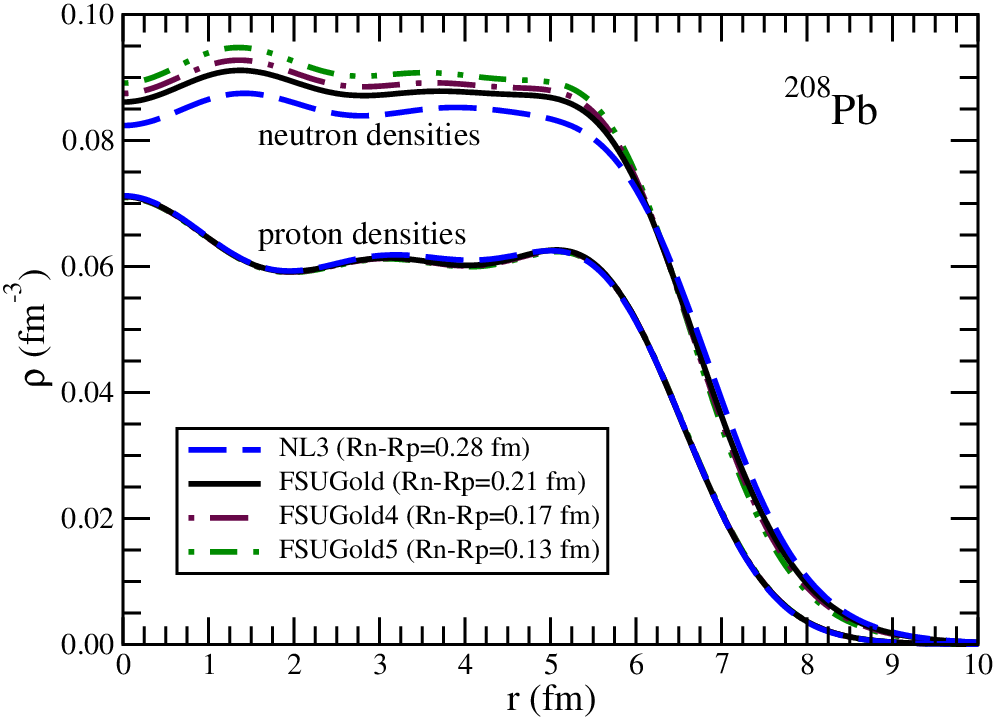}
\caption{Ground-state proton and neutron densities for ${}^{208}$Pb
         predicted by the various models described in the text.}
\label{Fig1}
\end{figure}

The model parameters used to compute the ground-state densities are
listed in Table~\ref{Table0}. Both NL3~\cite{Lal97_PRC55} and
FSUGold~\cite{Tod05_PRLxx} are accurately calibrated models obtained
from a fit to ground-state properties of finite nuclei.  The remaining
two sets, FSUGold4 and FSUGold5, are obtained from FSUGold by
adjusting the isovector parameters of the model ($g_{\rho}$ and
$\Lambda_{\rm v}$) as prescribed in
Ref.~\cite{Hor01_PRL86,Hor01_PRC64}.  This simple prescription enables
one to reduce the neutron skin in ${}^{208}$Pb without affecting the
properties of symmetry nuclear matter. Note, however, that a slight
reduction of the sigma-meson mass is also necessary to maintain the
charge radius of ${}^{208}$Pb within a fraction of one percent of its
experimental value.  In summary, we have used accurately calibrated
relativistic mean-field models to compute the ground-state densities
of ${}^{208}$Pb. As the neutron radius in ${}^{208}$Pb is not tightly
constrained by existent ground-state observables, a few of the
parameters of the model are adjusted to produce a wide range of values
for its nucleon skin, while maintaining the charge radius fixed at its
experimental value (see Table~\ref{Table1}). For a fairly
comprehensive recent review on how the ground-state densities are
computed see Ref.~\cite{Tod03_PRC67}. The various neutron and proton
densities generated by these models are displayed in Fig.~\ref{Fig1}.

  \begin{table}
  \begin{tabular}{|c|c|c|c|c|c|}
    \hline
     Observable & Experiment & NL3 & FSUGold 
                       & FSUGold4 & FSUGold5 \\
    \hline
     $B/A$~(MeV)            & $7.87$ & $7.88$ & 
                              $7.89$ & $8.10$ & $8.42$ \\
     $R_{\rm ch}$~(fm)      & $5.50$ & $5.51$ & 
                              $5.52$ & $5.52$ & $5.52$ \\
     $R_{n}\!-\!R_{p}$~(fm) &  --- & $0.28$ & 
                              $0.21$ & $0.17$ & $0.13$ \\
    \hline
  \end{tabular}
 \caption{Comparison between theory and experiment of the binding 
          energy per nucleon and the charge radii of ${}^{208}$Pb. 
          Theoretical predictions are also displayed for the 
	  neutron skin.}
  \label{Table1}
 \end{table}

\subsection{Scattering Formalism} 
\label{sec:scattering}

Having generated ground-state densities that will serve as input for
the optical potential, we present here a brief review of the
scattering formalism employed in this work. For some excellent reviews
of this very well developed field see, for example,
Ref.~\cite{Ray92_PR212} and references contained therein.

Assuming invariance under rotation, parity, and time reversal, the
model independent scattering amplitude for the elastic scattering
of a nucleon from a spinless target may be written in terms of two 
amplitudes as follows:
\begin{equation}
  \hat{{\cal F}}({\bf k},{\bf k}')=
   F_{0}(q,T_{\rm lab})+F_{1}(q,T_{\rm lab})
   \mbox{\boldmath$\sigma$}\cdot{\hat{\bf n}}\;,
 \label{ScattAmpl}
\end{equation}
where $\mbox{\boldmath$\sigma$}$ is the spin operator of the
projectile and 
${\hat{\bf n}}\!=\!\hat{{\bf q}}\!\times\!\hat{{\bf K}}$
is a unit vector perpendicular to the scattering plane. Here ${\bf q}$
and ${\bf K}$ denote the momentum transfer and average momentum of the
collision in the center-of-momentum frame, respectively.  That is,
\begin{equation}
 {\bf q}={\bf k}-{\bf k}' \quad{\rm and}\quad
 {\bf K}=({\bf k}+{\bf k}')/2 \;.
 \label{qandK}
\end{equation}
Note that for the kinematical dependence of both, the spin-independent
($F_{0}$) and spin-dependent ($F_{1}$), amplitudes we have selected
the kinetic energy of the projectile in the laboratory frame $T_{\rm
lab}$ and the magnitude of the momentum transfer in the $NA$
center-of-momentum frame $q\!=\!2k\sin(\theta/2)$. Other choices are
possible, yet the eikonal framework (discussed below) makes clear the
advantage of using $q$.

As the $NA$ scattering amplitude is written in terms of two
independent complex functions, three independent observables are
sufficient to fully characterize the reaction. It is customary to
select the differential cross section ($d\sigma/d\Omega$), the
analyzing power ($A_{y}$), and the spin-rotation function ($Q$) as the
three independent observables. They are given by the following
expressions:
\begin{subequations}
 \begin{equation}
  \frac{d\sigma}{d\Omega}=|F_{0}|^{2}+|F_{1}|^{2} \;,
  \label{dsigma}
 \end{equation}
 \begin{equation}
  A_{y}=2{\it Re}\left(
  \frac{F_{0}^{*}F_{1}}{|F_{0}|^{2}+|F_{1}|^{2}}\right) \;,
  \label{Ay}
 \end{equation}
 \begin{equation}
  Q=2{\it Im}\left(
  \frac{F_{0}^{*}F_{1}}{|F_{0}|^{2}+|F_{1}|^{2}}\right) \;.
  \label{Q}
 \end{equation}
\end{subequations}
As our main purpose is the extraction of the neutron radius of
${}^{208}$Pb, we limit ourselves, as other have done before us,
to study the differential cross section. 

\subsection{Non-relativistic Local KMT Optical Potential}
\label{sec:KMT}

The theoretical underpinning of most approximations to the elastic
scattering of fast nucleons from nuclei is contained in the seminal
work of Kerman, McManus, and Thaler (KMT)~\cite{Ker59_Ann08}. While 
it is not our intention to provide a comprehensive review of the
formalism (as an extensive literature is available) it is convenient
to highlight a few aspects of the present approach. 

To start, the nucleon-nucleus elastic scattering amplitude is written 
in terms of the ``one-body'' optical potential as follows:
\begin{equation}
  \hat{{\cal F}}({\bf k},{\bf k}')=-\frac{m}{2\pi}
  \hat{{\cal T}}({\bf k},{\bf k}')=-\frac{m}{2\pi}
  \langle{\bf k}'|V_{\rm opt}|{\bf k}+\rangle\;,
 \label{ScattAmpl2}
\end{equation}
where $m$ is the nucleon mass and $|{\bf k}+\rangle$ denotes an
``$+i\epsilon$'' nucleon distorted wave, namely, a plane wave of
momentum ${\bf k}$ in the infinite past. The on-shell microscopic KMT
optical potential $V_{\rm opt}\!=\!V_{\rm KMT}$ for the scattering of
a nucleon from a target nucleus with baryon number $A$ is given by
\begin{equation}
  V_{\rm KMT}({\bf q})= \left(\frac{A-1}{A}\right) C(T_{\rm lab}) 
             \sum_{i=p,n}{\widehat t}_{0i}\left(q,T_{\rm lab}\right) 
             \rho_{i}(q) \;.
 \label{OptKMT}
\end{equation}
Note that for the scattering of medium-energy nucleons the impulse
approximation is invoked, namely, the KMT optical potential contains
the free $NN$ t-matrix ($\widehat{t}_{0i}$) between the projectile and
the $i$-th target nucleon.  Further, $\rho_{i}(q)$ denotes proton and
neutron form factors obtained from the Fourier transform of their
respective ground-state densities. Finally, the whole expression is
multiplied by the following kinematical factor:
\begin{equation}
  C(T_{\rm lab}) \equiv 
   \left(\frac{2}{E_{p}}\right)_{NN}
   \left(\frac{E_{p}E_{t}}{E_{p}+E_{t}}\right)_{NA} \;,
 \label{kinconst}
\end{equation}
which preserves the unitarity of the nucleon-nucleon amplitude as 
it is transformed into the nucleon-nucleus CM
frame~\cite{Pic84_PRC30}.  Note that $E_{p}$($E_{t}$) is the
projectile(target) energy evaluated in the center-of-momentum frame of
either the $NN$ or $NA$ system. Although the formalism presented here
is intrinsically non-relativistic, all kinematical variables are
evaluated using relativistic kinematics.  For nucleon scattering from
a spin-saturated nucleus (such as ${}^{208}$Pb) the only pieces of the
elementary $NN$ amplitude that contribute to the nucleon-nucleus
elastic reaction are the central and spin-orbit pieces. That is, for
the scattering from a spin-saturated nucleus one may write
\begin{equation}
{\widehat f}_{0i}(q,T_{\rm lab})=-\frac{m}{2\pi}
{\widehat t}_{0i}(q,T_{\rm lab}) 
\rightarrow 
   A(q,T_{\rm lab})+iC(q,T_{\rm lab})\,
   \mbox{\boldmath$\sigma$}\cdot{\hat{\bf n}} \;,
 \label{Wolfenstein}
\end{equation}
where in the impulse approximation both of the ``Wolfenstein'' amplitudes 
($A$ and $C$) are directly inferred from free $NN$ scattering data. 

Having generated both nuclear form factors and the relevant components of 
the free $NN$ t-matrix, one may now proceed to compute the elastic scattering 
cross section. This is implemented by solving the resulting Lippmann-Schwinger 
equation for the nucleon-nucleus t-matrix,
\begin{equation}
  \widehat{{\cal T}}_{KMT}({\bf k},{\bf k}')=
        \langle{\bf k}'|V_{\rm KMT}|{\bf k}+\rangle
       =\langle{\bf k}'|V_{\rm KMT}|{\bf k} \rangle
       +\langle{\bf k}'|V_{\rm KMT}G_{0}^{(+)}V_{\rm KMT}
                                   |{\bf k}+\rangle\;,
 \label{ScattAmplKMT}
\end{equation}
directly in momentum space through standard numerical
techniques~\cite{Pic84_PRC30}. Note that $G_{0}^{(+)}$ denotes the
free projectile propagator. It is important to mention that within the
scope of the impulse approximation---and with ground-state form
factors generated from an accurately calibrated
parametrization---there are no adjustable parameters in the
calculation.

\subsection{Eikonal Approximation}
\label{eikonal}

The eikonal approximation provides a simple, physically illuminating
picture of the scattering process at intermediate energies. In what
follows we quote freely from the insightful review article by
Amado~\cite{Ama85_ANP15}. In the eikonal approximation the elastic
scattering amplitude is given in terms of an integral over impact
parameter that effectively sums up the myriad of partial waves
required at these intermediate energies. Moreover, at these energies
the integrand oscillates rapidly making the scattering amplitude
ideal for evaluation by the method of stationary phase.  Through this
analytic approach, the main features of the elastic $NA$ reaction
emerge in a clean and transparent way~\cite{Ama80_PRC21}. Indeed, the
diffractive oscillations are controlled by the nuclear size, the
exponential falloff by the diffuse nuclear surface, and the filling of
the minima by Coulomb effects, the real part of the $NN$ t-matrix, and
spin effects. It is only this filling that depends on the underlying
dynamics. Two of the main features of the diffractive cross
section---oscillations and exponential decay---are completely
determined by the nuclear geometry.

The spin-independent part of the scattering amplitude may be written
in the eikonal approximation as an integral over impact parameter. 
That is,
\begin{equation}
  F_{0}(q,T_{\rm lab})=\frac{iK}{2\pi}
                 \int d^{2}b\,e^{i{\bf q}\cdot{\bf b}}
                 \left(1-e^{-\chi({\bf b})}\right)\;,
 \label{EikonalAmpl0}
\end{equation}
where the profile function $\chi({\bf b})$, still an operator in the
spin space of the nucleon, is given by the following expression:
\begin{equation}
 \chi({\bf b})=i\frac{m}{K}\int_{-\infty}^{\infty}
               dz V_{\rm opt}({\bf r}) \;,
 \label{Profile}
\end{equation}
where $V_{\rm opt}({\bf r})$ is the Fourier transform of the (assumed 
local) optical potential
\begin{equation}
  V_{\rm opt}({\bf r})= \int \frac{d^{3}q}{(2\pi)^{3}}
           e^{i{\bf q}\cdot{\bf r}}\,V_{\rm opt}({\bf q})\;.
 \label{Vopt0}
\end{equation}
In the KMT approximation the optical potential $V_{\rm opt}({\bf q})$
is given by the form displayed in Eq.~(\ref{OptKMT}). 

While in Sec.~\ref{sec:results} results will be presented following
the KMT formalism without any further approximation, the eikonal
results include additional approximations so that contact may be
established with the analytic approach developed by Amado and
collaborators~\cite{Ama80_PRC21,Ama85_ANP15}. One of these
approximations emerges from comparing the range of the nuclear form
factor relative to that of the central and spin-orbit pieces of the
$NN$ interaction (Eq.~\ref{Wolfenstein}). For a spin-saturated
nucleus, the long-range pion potential---which is strongly
spin-dependent---does not contribute to leading order. This suggests
that both the central and spin-orbit pieces of the $NN$ interaction
are short ranged.  Alternatively, in a meson-exchange picture, only
multiple-pion exchanges (the so-called ``sigma'' meson) and
short-range mesons (like the omega and the rho) are allowed to
contribute to the elastic reaction. Thus, the relevant range of the
$NN$ interaction (of a fraction of a fm) is much shorter than the
nuclear radius (of several fm). This suggests that the KMT optical
potential may be written as
\begin{equation}
  V_{\rm KMT}({\bf r}) \approx 
  \left(\frac{A-1}{A}\right) C(T_{\rm lab})
  \left[
   {\widehat t}_{0p}\left(q\!\equiv\!0,T_{\rm lab}\right)\rho_{p}(r)+
   {\widehat t}_{0n}\left(q\!\equiv\!0,T_{\rm lab}\right)\rho_{n}(r)
  \right] \;.
 \label{Vopt1}
\end{equation}
The mismatch in ranges implies that only the forward-angle
part of the free $NN$ scattering amplitude is required.  Further, as
the only remaining radial dependence is contained in the ground-state
densities, the optical potential becomes spherically symmetric. This
yields the following form for the scattering amplitude:
\begin{equation}
  F_{0}(q,T_{\rm lab})\approx\frac{iK}{2\pi}
                 \int_{0}^{\infty} bdb J_{0}(qb)
                 \left(1-e^{-\chi(b)}\right)\;,
 \label{EikonalAmpl1}
\end{equation}
where $J_{0}$ denotes the Bessel function of order $n\!=\!0$. As the 
differential cross section---the sole elastic observable
presented in this work---involves the incoherent sum of a large
spin-independent amplitude and a small spin-dependent one [see
Eq.~(\ref{dsigma})], the spin-dependent amplitude will be neglected 
in the eikonal formalism henceforth.

\section{Results} 
\label{sec:results}

In this section differential cross sections are presented for the
elastic scattering of $T_{\rm lab}\!=\!500$~MeV~\cite{Hut88_NPA483}
and $T_{\rm lab}\!=\!800$~MeV~\cite{Hof80_PRC21} protons from
${}^{208}$Pb.  This section is itself divided into three
subsections. In the first subsection results are presented according
to the formalism outlined in Sec.~\ref{sec:KMT}. Having defined the
KMT form of the optical potential in Eq.~(\ref{OptKMT}), no further
approximations are made. Moreover, as the free $NN$ t-matrix is used
as input and the ground-state nuclear densities are generated from
accurately calibrated models, there is no room left for any adjustable
parameters.  The free NN t-matrices employed in our calculations are
obtained from a meson-exchange model including Delta-isobar degrees of
freedom to account for pion-production above the pion threshold. The
model is based on the work of Ref.~\cite{Els88_PRC38} and has recently
been revised and updated to fit presently available proton-proton and
neutron-proton data up to about 1.2 GeV~\cite{Sch04_Thesis}.  Many
other choices for the $NN$ t-matrix are possible provided they yield a
faithful representation of the experimental data.  As we shall see,
the KMT results compare very favorable to experimental data at both
energies.  In the second subsection results are presented in the
simpler eikonal approach and are compared against the previously
obtained KMT predictions.  The goal of this comparison is to use the
analytic power of the eikonal approach to elucidate the main features
of the data.  In particular, we aim to understand the central result
of our work: the insensitivity of the differential cross section to
our input neutron densities. We reserve the last subsection to address
this important point.

\subsection{Non-relativistic KMT Results} 
\label{sec:KMTresults}

\begin{figure}[ht]
\vspace{0.50in}
\includegraphics[width=5in,angle=0]{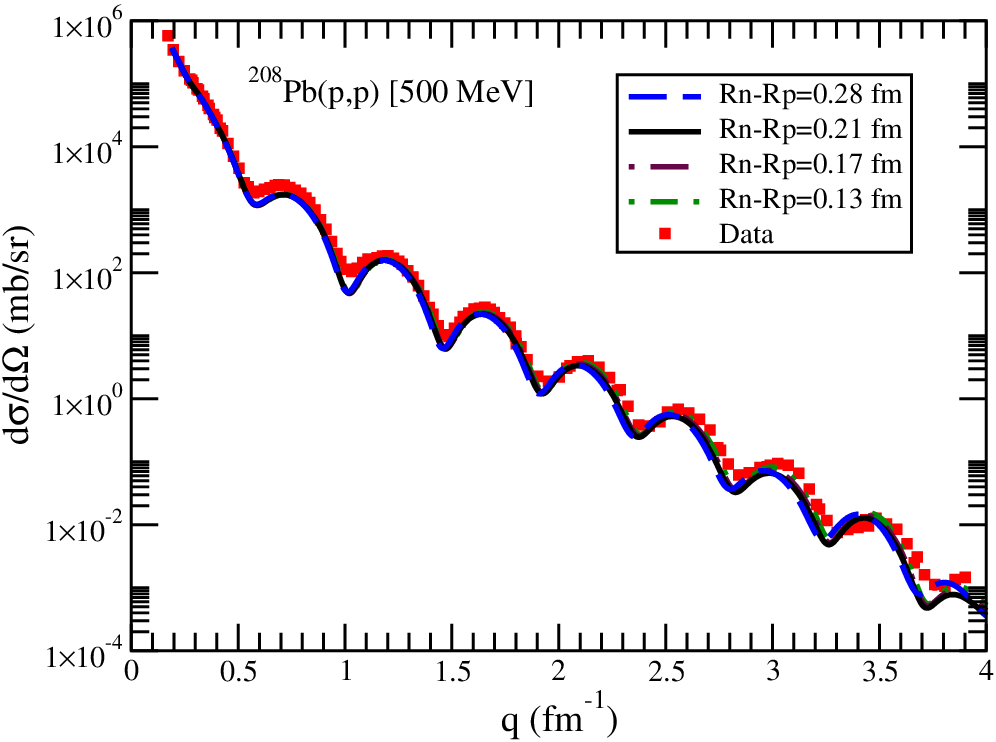}
\caption{Elastic scattering cross section of $T_{\rm lab}\!=\!500$~MeV 
         protons from ${}^{208}$Pb as a function of the momentum transfer
         to the nucleus. All theoretical models use the KMT optical 
         potential of Eq.~(\ref{OptKMT}) with ground-state densities 
         having values for the neutron skin of ${}^{208}$Pb as displayed 
         in the legend. The experimental data is from 
         Ref.~\cite{Hut88_NPA483}.}
\label{Fig2}
\end{figure}

\begin{figure}[ht]
\vspace{0.50in}
\includegraphics[width=5in,angle=0]{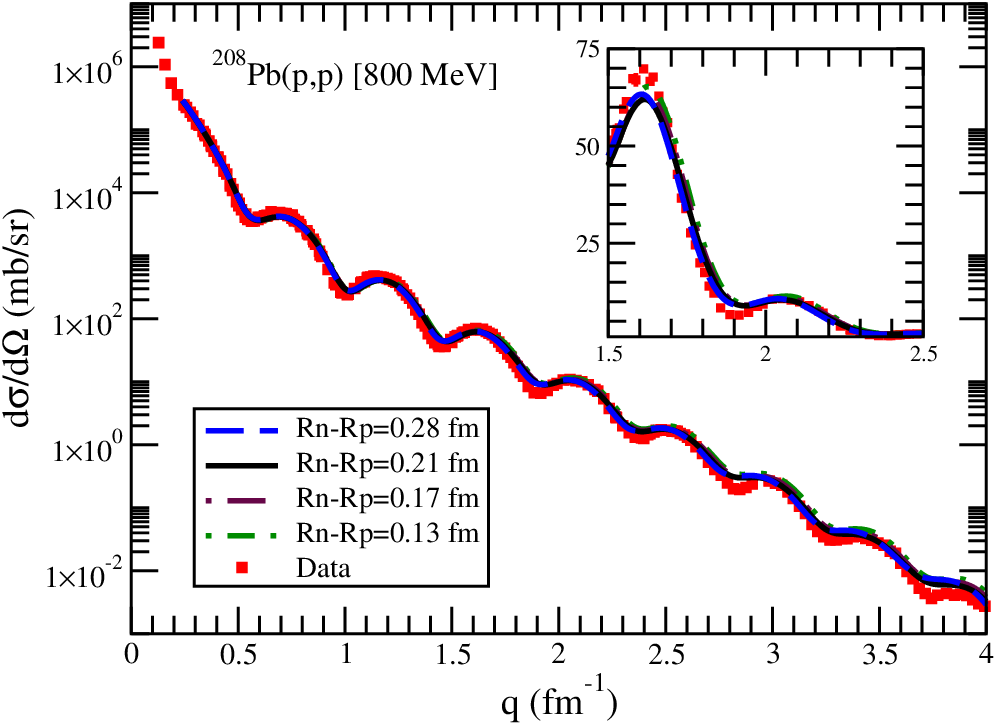}
\caption{Elastic scattering cross section of $T_{\rm lab}\!=\!800$~MeV 
         protons from ${}^{208}$Pb as a function of the momentum transfer
         to the nucleus. All theoretical models use the KMT optical 
         potential of Eq.~(\ref{OptKMT}) with ground-state densities 
         having values for the neutron skin of ${}^{208}$Pb as displayed 
         in the legend. The inset shows the differential cross section
         on a linear scale over a limited range of momentum transfers.
	 The experimental data is from Ref.~\cite{Hof80_PRC21}.}
\label{Fig3}
\end{figure}

The results presented in this section are based on the KMT formalism
reviewed in Sec.~\ref{sec:KMT}. The differential cross section for the
elastic scattering of $T_{\rm lab}\!=\!500$~MeV and $T_{\rm
lab}\!=\!800$~MeV protons from ${}^{208}$Pb is displayed in
Figs.~\ref{Fig2} and~\ref{Fig3}, respectively. The four curves
displayed in the figure use the same identical input from the $NN$
sector and differ only in the choice of ground-state densities as
described in Sec.~\ref{sec:structure}; the predicted neutron-skin
thickness for the various models is indicated in the legend. The
experimental data is from Refs.~\cite{Hut88_NPA483}
and~\cite{Hof80_PRC21}, respectively.

The description of the experimental data---with any one of the four
nuclear-structure models---is highly satisfactory. Indeed, the level
of success is comparable to that of recent studies that aim to
constrain the neutron radius of
${}^{208}$Pb~\cite{Kar02_PRC65,Cla03_PRC67,Dup05_xxx}. Further, while
theory appears to have difficulty in reproducing data at large
momentum transfers ($q\!\agt\!3.5$~fm), this happens only after the
cross section has dropped by about 8 orders of magnitude. Finally, the
observed differences among the four nuclear-structure models is so
minute that no clear preference for any of the models emerges. Note
that this remains true even if one plots the data in a 
linear scale, as shown on the inset of Fig.~\ref{Fig3}.

\subsection{Non-relativistic Eikonal Results} 
\label{sec:Eikonalresults}

In this section we are willing to sacrifice some degree of accuracy in
favor of analytic insight~\cite{Ama80_PRC21,Ama85_ANP15}. It is not
the aim of this section to replace the highly accurate description
achieved above, but rather to complement it with what we hope will be
some useful insights. The eikonal calculation reported here differs in
three important ways relative to the more accurate KMT approach. First,
the mismatch between the small $NN$ range and the large nuclear range
enables one to write the optical potential in the form given in
Eq.~(\ref{Vopt1}). Second, the small spin-dependent component of the
scattering amplitude ($F_{1}$) is neglected in the description of
the incoherent cross section [see Eq.~(\ref{dsigma})]. Finally, no
long-range Coulomb effects are included. In particular, the omission
of Coulomb effects will become clearly distinctive in the diffractive
minima of the cross section.

\begin{figure}[ht]
\vspace{0.50in}
\includegraphics[width=5in,angle=0]{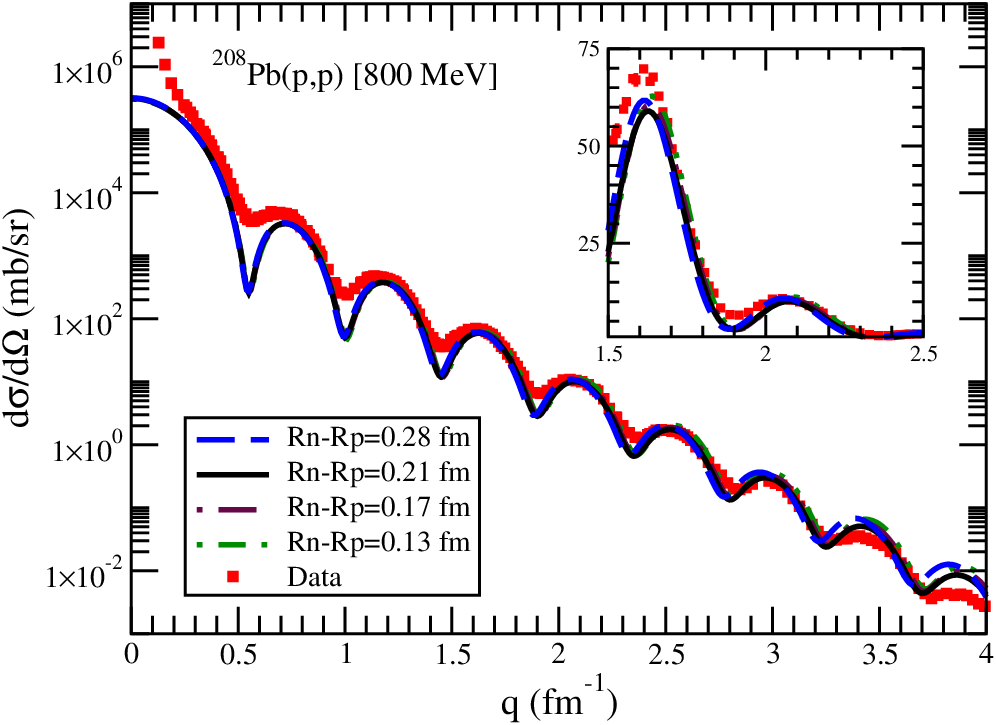}
\caption{Elastic scattering cross section of $T_{\rm lab}\!=\!800$~MeV 
         protons from ${}^{208}$Pb as a function of the momentum transfer
         to the nucleus. All theoretical models use the simple optical 
         potential of Eq.~(\ref{Vopt1}) with ground-state densities 
         having values for the neutron skin of ${}^{208}$Pb as displayed 
         in the legend. The inset shows the differential cross section
         on a linear scale over a limited range of momentum transfers.
	 The experimental data is from Ref.~\cite{Hof80_PRC21}.}
\label{Fig4}
\end{figure}

The eikonal results for the differential cross section at $T_{\rm
lab}\!=\!800$~MeV are displayed in Fig.~\ref{Fig4}, with the various
predictions labeled as in Figs.~\ref{Fig2} and~\ref{Fig3}.  Evidently,
there is a significant loss in quality relative to the full KMT
calculation. The cross section is underestimated at small angles,
overestimated at large angles, and the predicted minima are too
deep. Remarkably, however, the diffractive oscillations and the
exponential decay of the cross section are well reproduced. It is the
aim of the next section to elucidate these features. Note in closing
that as in the KMT case, there is no discernible difference between
the models, although some slight separation starts to appear at large
momentum transfers ($q\!\agt\!3.5$~fm).

\subsection{Discussion} 
\label{sec:discussion}

In this section we resort to analytic insights to elucidate the main
features of the elastic cross section, namely, exponentially modulated
diffractive oscillations with filled in
minima~\cite{Ama80_PRC21,Ama85_ANP15}. In comparing the eikonal
results to the experimental data (Fig.~\ref{Fig4}), the most glaring
deficiency of the model appears in minima that are too deep. Yet the
rapid diffractive oscillations and the exponential falloff are
relatively well described. The filling of the minima is strongly
sensitive to the underlying $NN$ dynamics, specifically to the real
part of the elementary $NN$ t-matrix and the Coulomb
interaction. While the former is included in our eikonal calculations
[see Eq.~\ref{Vopt1}], the latter is not.  Although scattering from a
heavy nucleus is strongly sensitive to Coulomb effects, we have
decided to neglect them in the eikonal approach (recall, however, that
the Coulomb interaction has been fully incorporated in our KMT
results~\cite{Chi91_PRC44}). Indeed, rather than adding to the eikonal
formalism, we will be removing from it---so we can dissect the various
components driving the elastic reaction.

\begin{figure}[ht]
\vspace{0.50in}
\includegraphics[width=5in,angle=0]{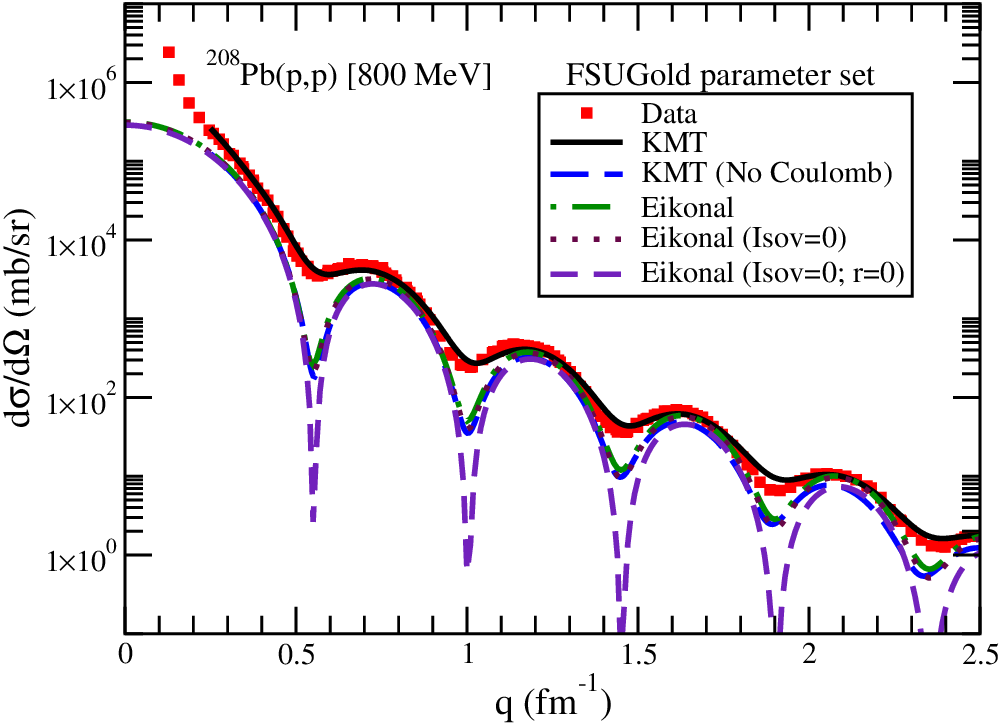}
\caption{Elastic scattering cross section of $T_{\rm lab}\!=\!800$~MeV 
         protons from ${}^{208}$Pb as a function of the momentum transfer
         to the nucleus. The various theoretical approximations are 
	 discussed in the text. All theoretical calculation have been 
	 performed using ground-state densities from the FSUGold parameter 
         set. The experimental data is from Ref.~\cite{Hof80_PRC21}.}
\label{Fig5}
\end{figure}

We start from Eq.~(\ref{Vopt1}) and rewrite the optical potential 
in an isoscalar-isovector basis rather than a proton-neutron one. 
That is,
\begin{equation}
  V_{\rm KMT}({\bf r}) \approx 
  \left(\frac{A-1}{A}\right) C(T_{\rm lab})
  \left[
   {\widehat t}_{0}\left(q\!\equiv\!0,T_{\rm lab}\right)\rho_{0}(r)+
   {\widehat t}_{1}\left(q\!\equiv\!0,T_{\rm lab}\right)\rho_{1}(r)
  \right] \;,
 \label{Vopt2}
\end{equation}
where 
$\widehat{t}_{0/1}\!=\!({\widehat t}_{0p}\pm{\widehat t}_{0n})/2$ are
the isoscalar/isovector components of the $NN$ t-matrix at forward
angles, $\rho_{0}\!=\!\rho_{p}\!+\!\rho_{n}$ is the isoscalar (or
matter) density, and $\rho_{1}\!=\!\rho_{p}\!-\!\rho_{n}$ is the
isovector density.  There are two factors that make the isovector
contribution to the cross section small. First, the isovector density
is reduced relative to the isoscalar density by roughly a factor of
$(N\!-\!Z)/(N\!+\!Z)\!\approx\!0.21$. Second, at the energies at which
the impulse approximation is valid, the $NN$ amplitude is
predominantly isoscalar. To illustrate this point we list in
Table~\ref{Table2} the $pp$, $pn$, isoscalar, and isovector scattering
amplitudes in the forward direction as a function of the projectile
kinetic energy using the $NN$ amplitudes of Ref.~\cite{Sch04_Thesis}.
These numbers suggest that relative to the dominant
isoscalar contribution, the isovector contribution provides at most a
5\% correction to the $NA$ scattering amplitude. This point will later
be validated by our numerical results (see
Fig.~\ref{Fig5}). Neglecting the isovector contribution yields an
eikonal scattering amplitude of the following simple form:
\begin{equation}
  F_{0}(q,T_{\rm lab})\approx\frac{iK}{2\pi}
                 \int_{0}^{\infty} bdb J_{0}(qb)
                 \left(1-e^{-\gamma\,t(b)}\right)\;,
 \label{EikonalAmpl2}
\end{equation}
where the profile function of Eq.~(\ref{Profile}) now becomes
\begin{equation}
  \chi(b) \approx \gamma\,t(b)
          = \frac{\sigma_{\rm tot}}{2}(1-ir)
            \int_{-\infty}^{\infty}dz\,\rho_{0}(r) \;.
 \label{Profile1}
\end{equation}
Here $\sigma_{\rm tot}$ is the total $NN$ cross section and $r$ is the
ratio of real to imaginary part of the $NN$ forward amplitude. These 
two equations represent the starting point of the analytic study of 
Ref.~\cite{Ama80_PRC21}.

\begin{table*}
\begin{tabular}{|l|c|c|c|c|c|}
 \hline
  $T_{\rm lab}$ & $A_{pp}$ & $A_{pn}$ & $A_{\rm isos}$ 
    & $A_{\rm isov}$ & $|A_{\rm isov}|/|A_{\rm isos}|$ \\
 \hline
  400 & $(+0.1949,0.3659)$ & $(+0.0874, 0.5441)$ 
      & $(+0.1412,0.4550)$ & $(+0.0538,-0.0891)$ 
      & $0.2185$ \\
  500 & $(+0.0983,0.4367)$ & $(-0.0364, 0.5733)$ 
      & $(+0.0310,0.5050)$ & $(+0.0674,-0.0683)$ 
      & $0.1897$ \\
  600 & $(+0.0111,0.5098)$ & $(-0.1446, 0.6051)$ 
      & $(-0.0668,0.5575)$ & $(+0.0779,-0.0477)$ 
      & $0.1627$ \\
  700 & $(-0.0744,0.5885)$ & $(-0.2434, 0.6430)$ 
      & $(-0.1589,0.6158)$ & $(+0.0845,-0.0273)$ 
      & $0.1396$ \\
  800 & $(-0.1624,0.6668)$ & $(-0.3385, 0.6819)$ 
      & $(-0.2505,0.6744)$ & $(+0.0881,-0.0076)$ 
      & $0.1229$ \\
 \hline
\end{tabular}
\caption{Central component of the $NN$ amplitude (in fm) in the 
         forward direction as a function of the laboratory energy 
         (in MeV) of the projectile.}
\label{Table2}
\end{table*}

To separate $NN$ dynamics from nuclear geometry we display in
Fig.~\ref{Fig5} various calculations that include several levels of
approximation. In all these calculations the common denominator is the
nuclear density. Only nuclear densities generated from the FSUGold
parameter set are included in this illustration. The curve with the
very deep minima (short dashes) is generated from Eq.~(\ref{Vopt2}) by
setting both the isovector term (second term in the expression between
brackets) and the real part of the $NN$ amplitude [$r$ in
Eq.~(\ref{Profile1})] to zero. The same exact calculation but with $r$
restored to its physical value yields the curve displayed by the
dotted line.  As alluded in Refs.~\cite{Ama80_PRC21,Ama85_ANP15}, we
confirm that the real part of the elementary amplitude plays an
essential role in filling in the minima. Next we move to the curve
represented by the dot-dashed line. This is identical to the one
displayed by the black solid line in Fig.~\ref{Fig4} and differs from
the previous (dotted) one only in that the isovector term has been
restored. The imperceptible difference among these two calculations
suggests that at the energies considered here the scattering amplitude
is strongly dominated by the isoscalar contribution.  Next we compare
against KMT results. Recall that the KMT scattering amplitude is
computed from the optical potential of Eq.~(\ref{OptKMT}) without any
further approximation. However, in an effort to compare with our
eikonal results the Coulomb interaction has been momentarily turned
off from the KMT calculation (long dashed line). The difference
between this calculation and the previous two eikonal results is
minimal. The full KMT calculation with the Coulomb interaction
restored (solid line) fills in nicely the minima and provides a
faithful representation of the scattering data. Remarkably, the
various approximations considered in Fig.~\ref{Fig5} affect mostly the
minimum-filling of the cross section; the diffractive oscillations and
exponential falloff of the cross section are robust and, thus, fairly
insensitive to the various approximations.

In summary, the eikonal formalism reveals that the main features
displayed by the differential cross section are dominated by the
nuclear edge, with the radius of the nucleus controlling the
diffractive oscillations and the surface thickness the exponential
decay. Further, although the diffraction minima are well developed,
the minima are filled in---with the filling arising predominantly from
the real part of the $NN$ interaction and Coulomb effects. Finally, at
the medium energies considered in this work (and necessary for the
impulse approximation to be valid) the optical potential is 
strongly dominated by isoscalar effects.

Within the above context, the insensitivity of the elastic reaction 
to the nuclear radius should no longer come as a surprise. As argued
above, the diffractive oscillations in this medium-energy region are
determined by the matter radius $R_{0}$, which is given by
\begin{equation}
  R_{0} = \sqrt{\frac{Z}{A}R_{p}^{\,2}+\frac{N}{A}R_{n}^{\,2}}
    \approx R_{p}\left(1+\frac{N}{A}\frac{\Delta R}{R_{p}}\right)
    \approx \left[5.46+0.23\left(\frac{\Delta R}{0.21}\right)
           \right] {\rm fm} \;.
 \label{RMatter}
\end{equation}
This implies that the more than a factor of two difference in the
model dependence of the neutron skin ($\Delta
R\!=\!0.13\!-\!0.28$~fm), gets diluted into a mere 1.5\% difference in
the matter radius. It is unrealistic to expect that the theoretical
uncertainties in this purely hadronic reaction will ever be resolved 
to the required level of precision.

\section{Conclusions} 
\label{sec:conclusions}

We have examined the sensitivity of elastic proton scattering to the
neutron radius of ${}^{208}$Pb at projectile energies of $T_{\rm
lab}\!=\!500$ and $T_{\rm lab}\!=\!800$~MeV. This energy region was
selected to minimize the controversial uncertainties in the reaction
mechanism inherent to all processes involving strongly-interacting
probes. By assuming the validity of the impulse approximation, we
exclusively relied on free $NN$ amplitudes that have been determined
directly from experimental data. For the nuclear-structure input we
relied on two accurately calibrated parameter sets, namely,
NL3~\cite{Lal97_PRC55} and FSUGold~\cite{Tod05_PRLxx}. While both sets
reproduce binding energies and charge radii of a variety of
nuclei, they predict neutron skins in ${}^{208}$Pb that differ
considerably: $\Delta R\!=\!0.28$~fm for NL3 and $\Delta
R\!=\!0.21$~fm for FSUGold.  However, in an effort to expand this
range to the region suggested by some recent
analyzes~\cite{Kar02_PRC65,Dup05_xxx}, the isovector parameters of the
FSUGold interaction were tuned to produce two additional parameter
sets, one with $\Delta R\!=\!0.17$~fm and the other one with $\Delta
R\!=\!0.13$~fm. This tuning was done while maintaining the charge
radius of ${}^{208}$Pb within 0.5\% of its experimental value.

Calculation were performed in an impulse approximation with a KMT
optical potential of the ``$t\rho$'' form [see
Eq.~(\ref{OptKMT})]. The resulting Lippmann-Schwinger equation for the
$NA$ amplitude was then solved without resorting to any further
approximation. Results for the differential cross section at both
energies (500 and 800 MeV) were in good agreement with the
experimental data. As important, no clear separation among the
nuclear-structure models was observed. This insensitivity of the
elastic cross section to the neutron skin of ${}^{208}$Pb represents
the main conclusion of this work.

In an effort to understand our numerical results we invoked the
eikonal approximation, a formalism that has contributed important
insights to our physical understanding~\cite{Ama80_PRC21,Ama85_ANP15}.
First, we established that in the medium-energy region considered
here, the cross section is predominantly sensitive to the matter
density. Second, we have verified that while the filling of the
diffraction minima arises from a combination of the Coulomb force and
the real part of the $NN$ interaction, the diffractive oscillations
and its exponential falloff are controlled by the nuclear
geometry---with the radius controlling the former and the diffuseness
the latter. Since the diffractive oscillations of the cross section are
controlled by the matter radius [see Eq.~(\ref{RMatter})], the large
model dependence of the neutron skin got diluted into an unobservable
1.5\% difference in the matter radius.

In summary, we find the elastic proton reaction insensitive to the
neutron radius of ${}^{208}$Pb. This stands in contradiction to recent
results ~\cite{Kar02_PRC65,Cla03_PRC67,Dup05_xxx}. Thus, we close with a
few comments. In Ref.~\cite{Kar02_PRC65} a neutron skin in ${}^{208}$Pb
of $\Delta R\!\simeq\!0.17$~fm was suggested from an analysis that
included nuclear densities obtained from three different Skyrme
interactions and two harmonic-oscillator (HO) parameterizations.
While the data can easily discriminate between the Skyrme and HO
models, distinguishing among the three Skyrme interactions becomes
possible only at large angles for the largest energy considered
($T_{\rm lab}\!=\!200$~MeV). Unfortunately, in the large-angle region
where the model dependence is the largest, there also appears to be a
significant discrepancy between theory and experiment. 
In the relativistic approach of Ref.~\cite{Cla03_PRC67} an optimal set
of neutron-density parameters is obtained by means of a global fit to
a large body of scattering observables. From this analysis the neutron
skin of ${}^{208}$Pb was constrained to the range $\Delta
R\!\sim\!0.083\!-\!0.111$~fm.  However, what is sorely missing from
these two analyzes is a discussion on how much will the description of
the differential cross section suffer by including other realistic
neutron densities having similar surface properties but different
neutron radii.  For example, could the non-relativistic work of
Ref.~\cite{Kar02_PRC65} still provide a good description of the
scattering data by using neutron densities with the smaller neutron
radius of Ref.~\cite{Cla03_PRC67}?

A partial answer---in the affirmative---to this question has been
provided in Ref.~\cite{Dup05_xxx}.  By using the SkM* Skyrme
interaction of Ref.~\cite{Kar02_PRC65} together with a realistic RPA
interaction, an excellent description of the differential cross
section for the elastic scattering of $T_{\rm lab}\!=\!121$~MeV
protons from ${}^{208}$Pb was obtained. Thus, the range of acceptable
values of the neutron skin of ${}^{208}$Pb got extended to the region
$\Delta R\!\simeq\!0.13\!-\!0.17$~fm.  In this contribution we have
extended the range even further: to $\Delta
R\!\simeq\!0.13\!-\!0.28$~fm. Note, however, that we were unable to
rule out neutron skins outside this range.

Proton-induced reactions have evolved to a point that cross section
measurements over ten orders of magnitude are now routine. Even so, it
is unrealistic to expect them to evolve into precision tools, mainly
because of the large uncertainties inherent in any theoretical
description of strongly interacting reactions. Still, the decisive role
that these reactions will play in our understanding of novel
properties of exotic nuclei is without question. Yet for a clean and
model independent extraction of the neutron radius of ${}^{208}$Pb we
must resort to electroweak probes. At present, the Parity Radius
Experiment at the Jefferson Laboratory appears as the only alternative
for providing a unique experimental constraint on the thickness of
the neutron skin of a heavy nucleus.

\begin{acknowledgments}
We thank Brandon van der Ventel for many useful discussions and
Charlotte Elster for providing us with the $NN$-potential code 
and for a careful reading of the manuscript. S.P.W. gratefully 
acknowledges Kirby Kemper and the Physics Department at Florida 
State University for their hospitality during his sabbatical 
leave. This work was supported in part by DOE grant DE-FG05-92ER40750.
\end{acknowledgments}
\vfill\eject

\bibliography{ReferencesJP} 

\vfill\eject
\end{document}